\documentclass[prb,preprint,showkeys]{revtex4-1} 
\usepackage{float}
\usepackage{graphicx}
\usepackage{amsmath, amsfonts}
\usepackage{bm}
\usepackage{subcaption}
\usepackage{multirow}
\usepackage{xcolor}
\usepackage{hyperref}
\usepackage{amsmath}

\begin{document}

\title[mode=title]{Alternative winding patterns for twisted solenoid coils with improved characteristics for TRASE MRI}

\author{Nahid Ghomimolkar}
\affiliation{Department of Physics \& Astronomy, University of Manitoba, Winnipeg, Manitoba, Canada}

\author{Alexander E.\ Krosney}
\affiliation{Department of Physics, University of Winnipeg, Winnipeg, Manitoba, Canada}

\author{Christopher P.\ Bidinosti}
\email{c.bidinosti@uwinnipeg.ca}
\affiliation{Department of Physics, University of Winnipeg, Winnipeg, Manitoba, Canada}
\affiliation{Department of Physics \& Astronomy, University of Manitoba, Winnipeg, Manitoba, Canada}

\begin{abstract}
Transmit Array Spatial Encoding (TRASE) is an MRI technique in which spatial encoding is achieved using  phase gradients of the $B_1$ field. This approach offers potential advantages such as hardware simplicity and reduced acoustic noise. In this study, we present an assessment of various winding patterns for  twisted solenoid phase-gradient coils, including the simple twisted solenoid (with and without return wire), a double-wound twisted solenoid, and a discrete-loop twisted solenoid. We analyze the magnetic field uniformity and phase linearity of these configurations using Biot–Savart simulations. Our results show that both the double-wound and discrete loop designs  offer similar improvements over the simple twisted solenoid with return wire. The discrete loop pattern requires less wire than the double-wound version, making it the preferred option for practical coil construction and operation in a TRASE MRI system.
\end{abstract}

\keywords{Magnetic resonance imaging (MRI), Transmit array spatial encoding (TRASE), RF phase gradient coils, Twisted solenoid, Stream function}

\maketitle

\section{Introduction}

Transmit array spatial encoding (TRASE) provides an innovative replacement for traditional $B_0$  gradient-based spatial encoding in magnetic resonance imaging (MRI) technology. Originally proposed by Sharp and King \cite{Sharp}, TRASE makes use of phase gradients of the radiofrequency (RF) or $B_1$ field to achieve spatial encoding, thus avoiding the necessity of static-field gradient coils, along with their accompanying hardware and challenges such as acoustic noise.  Further work by these authors  demonstrated the potential of TRASE for high resolution imaging~\cite{Sharp2}. TRASE MRI continues to be an active area of research~\cite{stockmann, bohidar,  Sarty2021, Sarty2025, Sedlock2025}, including hardware development~\cite{Der2018, Purchase2019, nacher, Purchase2021} and the potential impact of RF pulse-related artefacts~\cite{bidinosti_artefacts,Bohidar2020,Bidinosti2022}.
As is the case with any MRI methodology,  RF coil design is critical to furthering the development of TRASE.  RF phase-gradient coils are still novel, however,  and while inspiration has been drawn from standard design methods -- including Maxwell-Helmholtz arrays~\cite{deng} and target field approaches~\cite{bellec138, bellec723, kumaragamage, nacher} -- TRASE has not yet had the benefit of the decades of RF coil research that has driven innovation in conventional MRI.

 A  more recent line of exploration for TRASE MRI is the twisted solenoid, which grew out of double-helical coil designs~\cite{goodzeit,meinke2003modulated,alonso,queval} and was originally employed by Sun et al.~\cite{sunjmr} 
  to generate linear RF phase profiles with enhanced efficiency for  MRI systems having transverse or vertical $B_0$ magnetic field geometries.
Subsequent innovations included  (i) geometrically-decoupled twisted solenoid coil sets to further improve phase linearity while reducing mutual inductance between  transmit elements~\cite{sunmrm}, and (ii) truncated designs to reduce the length of twisted solenoid coils while preserving phase gradient strength and field uniformity~\cite{sedlock2023truncated}.
 In this paper, we investigate alternative winding patterns for twisted solenoid coils, employing Biot-Savart computations to compare key characteristics of the $B_1$ field with previous designs~\cite{sunjmr,  sedlock2023truncated}.   We find that our double-wound and discrete loop designs offer improved magnitude and phase-gradient homogeneity, which are both desirable for TRASE MRI.  Neither design should increase the complexity of actual realizations of twisted solenoid coils, and the discrete loop pattern has the added advantage of requiring essentially the same length of wire as existing designs.  The process of coil truncation~\cite{sedlock2023truncated} readily extends to the discrete loop design as well.

\section{Parametric curves and winding patterns}
\label{paramcurves}

The Cartesian coordinates of the winding pattern at the heart of previous twisted solenoid designs~\cite{sunjmr,  sunmrm, sedlock2023truncated, Sedlock2025} are given by the parametric curve
\begin{equation}
P(\theta) =\left[a \cos(\theta), a \sin(\theta), A_n \sin(n \theta + \varphi) + \frac{h}{2\pi} \theta \right] \, ,
\label{parametric_curve1}
\end{equation}
with each quantity (and its associated SI unit) defined as follows~\cite{queval}:
\begin{itemize}
\item $\theta$ (rad) -- the azimuthal angle in cylindrical coordinates.
\item $a$ (m) -- the radius of the cylindrical former on which the coil is wound.
\item $n$ -- the multipole order of the coil. (For TRASE coils, $n=2$~\cite{sunjmr}.)
\item $A_n$ (m) --  the multipole modulation or twist amplitude. (For a uniform solenoid, $A_n$ = 0.)
\item $\varphi$ (rad) -- the winding shift. (This selects the phase gradient direction. See Table~2 in Ref.~\cite{sunjmr}.)
\item $h$ (m) -- the coil turn advance. (The coil pitch (turns/m) is $1/h$.)
\end{itemize}
In the limit of a tightly-wound coil with thin wires, this  can be represented by a surface current (on a cylinder of radius $a$) comprising three terms:
\begin{equation}
\bm{j}(\theta, z) = \frac{I}{ha} \left[ a \, \bm{\hat{\theta}} +  \left(A_n n \cos(n \theta + \varphi) + \frac{h}{2\pi} \right)  \bm{\hat{z}} \right] \, .
\label{jqueval} 
\end{equation}
The first term is a uniform azimuthal surface current; the second is a cosine-theta surface current along the axial direction; while the third term is a uniform axial surface current which, by Ampere's law, generates zero magnetic field inside the coil itself~\cite{queval}.

In light of this last observation,  we disregard the third term and seek to employ the apparatus of the stream function $\psi(\theta,z)$,   commonly used in coil design~\cite{brideson,peeren,bidinosti_saddle}. 
The usefulness of the stream function lies in the properties that (i) contours of $\psi$ represent the stream lines (or flow lines) of current on the cylindrical surface, and (ii) evenly spaced contours of $\psi$ contain equal integrated surface current.  As a result, evenly spaced contours of $\psi$ provide one directly with the discrete current paths (i.e.\ the wire winding patterns) that approximate~$\bm j$.

To within a constant, then, the stream function here can be written as
\begin{equation}
\psi(\theta, z) = K z -  K A_n \sin(n\theta + \varphi) \, ,
\label{stream_function}
\end{equation}
which gives the desired surface current components 
\begin{equation}
j_\theta = \frac{\partial \psi}{\partial z} = K \quad \mbox{and } \quad j_{z} = -\frac{1}{a} \frac{\partial \psi}{\partial \theta} = \frac{KA_n n}{a} \cos(n \theta + \varphi)  \, ,
\label{jnahid}
\end{equation}
where the constant $K$ is the magnitude of the uniform azimuthal surface current (equal to $I/h$ in the above example).  
The parametrized curve
\begin{equation}
P(\theta) = \left[a, \theta, A_n \sin(n\theta + \varphi) + z_i \right] \, 
\label{parametric_curve2}
\end{equation}
gives the cylindrical components of a contour of value $\psi =K z_i$ set by by the constant $z_i$. 
This can be verified by substituting $z=A_n \sin(n\theta + \varphi) + z_i $ into Eq.~\ref{stream_function}. (The subscript $i$ will serve as a loop index in due order.)
Converting to Cartesian components, the parametrized curve for contours of $\psi$ is
\begin{equation}
P(\theta) = \left[a \cos(\theta), a \sin(\theta),  A_n \sin(n\theta + \varphi) + z_i \right] \, .
\label{parametric_curve2b}
\end{equation}

\begin{figure}[htbp]
\centering
  \includegraphics[trim={0 0cm 0 0cm},clip,width=0.9\textwidth]{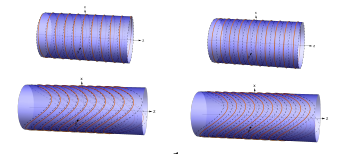}
  \caption{Conceptual examples of wire-wound approximations of the uniform (top row) and twisted (bottom row) solenoid. Left-side: single-wire pitch models via Eq.~\ref{parametric_curve1}. Right-side: discrete-loop models via Eq.~\ref{parametric_curve2b}. For all drawings, the red lines are the  wires, while the dashed lines indicate stream function contours (of Eq.~\ref{stream_function}) containing equal integrated surface current.  These contours are the same (left and right) for each of the two coil types (top with $A_n=0$ and bottom with $n=2$).}
    \label{3D_regular_twisted}
\end{figure}

To highlight the difference between the winding patterns resulting from Eq.~\ref{parametric_curve1} versus  Eq.~\ref{parametric_curve2b}, we present  two examples of each, side by side in Fig.~\ref{3D_regular_twisted}.  Uniform solenoids ($A_n=0$) are shown in the top row, while twisted solenoids of the type used in TRASE ($n=2$) are shown in the bottom row.  Evenly spaced contours of the stream function of Eq.~\ref{stream_function} for $A_n=0$ and $n=2$ are also shown.  All coils comprise ten turns.  As can be seen in the left-side examples, Eq.~\ref{parametric_curve1} results in coils wound of a single contiguous wire that, owing to the winding pitch, traverses a region of equal integrated surface current (of Eq.~\ref{jnahid}) following each turn.  Conversely, as seen in the right-side examples, Eq.~\ref{parametric_curve2} results in coils made of  discrete loops centred within each region of equal integrated surface current.

\begin{figure}[htbp]
\centering
  \includegraphics[width=0.7\textwidth]{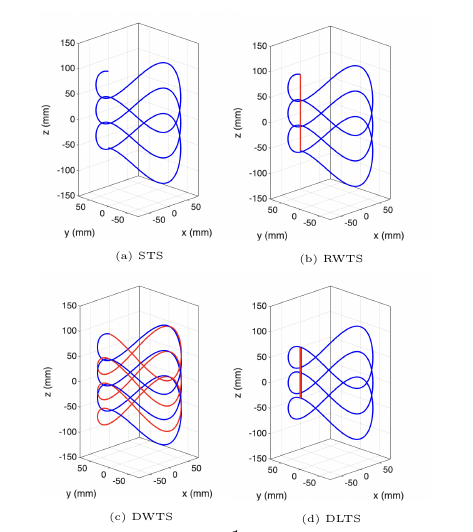}
  \caption{
  Examples of the four archetypal twisted solenoid winding patterns consider in this work: 
  (a) the simple twisted solenoid (STS), (b) the return-wire twisted solenoid (RWTS), (c) the double-wound twisted solenoid (DWTS), and (d) the discrete-loop twisted solenoid (DLTS).
  The red lines in the RWTS, DWTS and DLTS indicate return wires/windings.  
  The following parameters were used for all coils here: $a = 78$ mm, $A = 40$ mm, $\varphi= 90^\circ$, $h = 50$ mm, and $N = 3$.
  }
  \label{twisted_coils}
\end{figure}

In the remainder of the paper, we will explore and compare twisted solenoid coils based on these two different winding pattern concepts.   We will consider the four archetypal coil models presented in Fig.~\ref{twisted_coils}.  They are all defined by the parametric curve
\begin{equation}
\begin{bmatrix}
P_x(\theta) \\
P_y(\theta) \\
P_z(\theta)
\end{bmatrix}
=
\begin{bmatrix}
a \cos(\theta) \\
a \sin(\theta) \\
A \sin(2 \theta + \varphi) + H(\theta)
\end{bmatrix} \, ,
\label{parametric_curve3}
\end{equation}
where  
we have dropped the subscript on $A$ for simplicity and the function $H(\theta)$ is specific to each coil type as described immediately below.

For the simple twisted solenoid (STS), often contemplated in Biot-Savart simulations in Refs.~\cite{sunjmr,  sunmrm, sedlock2023truncated}, $H(\theta)=
h\theta / 2\pi $.  The parameter $\theta$  runs from $-N\pi$ to $N\pi$, where $N$ is the number of turns in the coil.  As shown in Fig.~\ref{twisted_coils}(a), there is no return wire associated with this model.  As such, it is  suitable for  exploring the broad characteristics of the twisted solenoid, but in general
should not be relied upon to provide detailed information  about any practical realization of such a coil.
To address this, we also consider the return-wire twisted solenoid (RWTS).  Here a single current segment at fixed azimuthal angle connects the last point of the last turn (i.e.\ $\theta = N \pi$) to the first point of the first turn (i.e.\ $\theta = -N \pi$) of an STS, thereby providing a closed current loop as shown Fig.~\ref{twisted_coils}(b).

The double-wound twisted solenoid (DWTS) is the first of our alternative approaches to the RWTS.  As shown in Fig.~\ref{twisted_coils}(c), it features a return STS to form a closed current loop and is thus the twisted-solenoid analogue of a two-layer uniform solenoid.  For the outgoing STS, $H(\theta)= h\theta / 2\pi $ with $\theta \in [-N\pi, N\pi]$ as above.  For the return STS, $H(\theta)= Nh - h\theta / 2\pi $ with $\theta \in [N\pi, 3N\pi]$, which connects the two sections at both ends.

The discrete-loop twisted solenoid (DLTS) is  our second alternative to the RWTS and is based on the stream function design method described above.  
Here, via Eq.~\ref{parametric_curve2b} with $z_i = hi$, $H(\theta)=  -h(N+1)/2  + hi $, where $i$ is the loop index that runs from $1$ to $N$ and  $\theta \in [-\pi, \pi]$ for each loop.  
For example, the first loop ($i=1$) follows the parametric curve of Eq.~\ref{parametric_curve3} with $H(\theta) = -hN/2 +h/2$, and lies at the midpoint between the contours of $\psi$ with $H(\theta) = - hN/2$ and $H(\theta) = - hN/2 + h$.
For the DLTS, then, $h$ is the distance between evenly spaced contours of $\psi$ (for fixed $\theta$).  It is equivalently the loop-to-loop spacing (for fixed $\theta$), and $1/h$ is therefore the axial loop density.  For direct comparison to 
the STS, RWTS or DWTS (Figs.~\ref{twisted_coils}(a)--(c)), one should choose $h$ to match the winding pitch of those models, as was done in Figs.~\ref{3D_regular_twisted} and~\ref{twisted_coils}.  In practice, the individual loops of the DLTS are connected in series, with a single return wire from the end of the $N^\mathrm{th}$ loop to the start of the $1^\mathrm{st}$ loop, as shown in Fig.~\ref{twisted_coils}(d).  The overlap of the connecting wires with the return wire nullifies their net magnetic field, and one need consider only  the closed current paths of the individual loops for Biot-Savart simulations~\cite{brideson,bidinosti_saddle}.

\section{$B_1$ field evaluation}
\label{sec:B1eval}

As was done by the authors of Refs.~\cite{sunjmr, sunmrm, sedlock2023truncated}, we use Biot-Savart simulations to evaluate the $B_1$ field produced by the various twisted solenoid designs explored here.   Consistent with their approach, we take  the $x$-axis to be the vertical direction (along $B_0$), and consider the $\bm B_1$ field comprising the transverse components $B_{1y}$ and $B_{1z}$  only.\footnote{The concomitant $B_{1x}$ field has negligible impact in standard realizations of TRASE~\cite{Sharp, deng, sunjmr, sunmrm, sedlock2023truncated} but may be important with very low $B_0$~\cite{bidinosti_artefacts}.}
We produce field maps and profiles of the magnitude of $\bm B_1$ 
\begin{equation}
|B_1| = \sqrt{B_{1y}^2 + B_{1z}^2} \, ,
\label{B1mag}
\end{equation}
the spatial phase of $\bm B_1$ 
\begin{equation}
\phi_B = \tan^{-1}\left(-\frac{B_{1y}}{B_{1z}}\right) \, ,
\label{B1phase}
\end{equation}
and the spatial gradient of $\phi_B$
\begin{equation}
G_r =  \frac{\partial \phi_B}{\partial r}\, ,
\label{B1phaseGrad}
\end{equation}
where $r$ is the phase gradient direction ($\pm x$, $\pm y$) set by the winding shift $\varphi$~\cite{sunjmr}.  Numerical simulation is  achieved by splitting the coils into many small straight-line current segments set by $1^\circ$ steps of the parameter $\theta$ in Eq.~\ref{parametric_curve3}, and subsequently calculating their contributions to the net magnetic field vector at a given point via in-house code (MATLAB) using the form of the Biot-Savart law for a filamentary segment from Ref.~\cite{hanson2002compact}.   The straight return wire of the RWTS coil is treated as a single current segment.  We use $18,000$ sample points for the 2D plots featured in the following sections, and 200 sample points for the 1D plots.  The phase gradient of Eq.~\ref{B1phaseGrad} is determined numerically using the built-in \textit{gradient} function in MATLAB.

\section{Results}
\subsection{Comparison of  the four archetypal twisted solenoid winding patterns }

The winding parameters for all coils in this study are $N=10$, $a=7.8$~cm, $A =5.5$~cm, $\varphi =0$, $h = 3.0$~cm.  This is a $G_x$ coil.
The parameters were taken from coil number \#1 in Table~3 of Ref.~\cite{sunjmr}, which was the top ranking design that came out of a parameter search study, subject to  selection criteria on the basis of $B_1$ field uniformity and phase gradient strength, followed by a ranking criterion based on a coil efficiency metric.  
A version of this coil was constructed and used along with a uniform saddle coil for a successful demonstration of 1D TRASE MRI~\cite{sunjmr}. Field mapping with a pick-up coil showed close agreement with Biot-Savart simulations~\cite{sunjmr}.

\begin{figure}[htbp]
  \includegraphics[width=\textwidth]{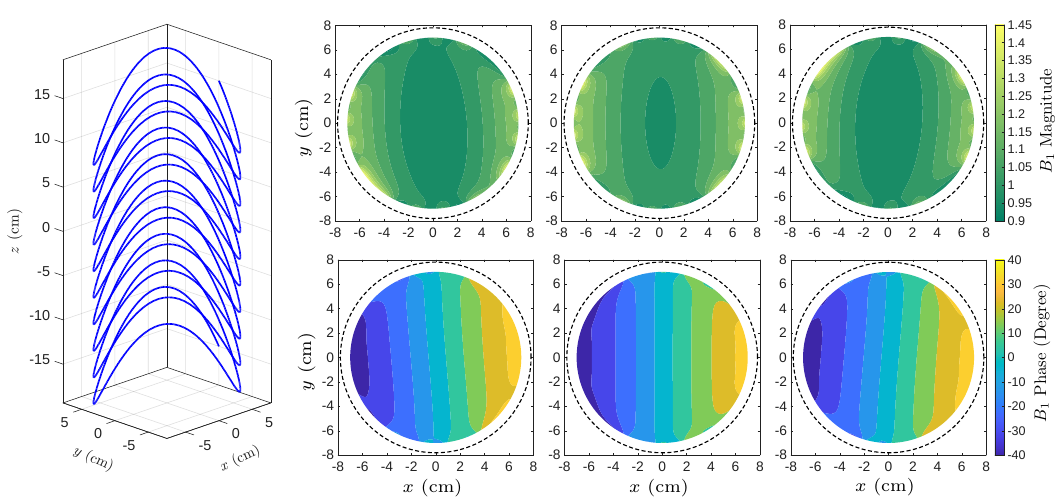}
  \captionof{figure}{Results for the simple twisted solenoid (STS). The left plot shows the STS winding pattern. The top row of contour plots on the right displays $B_1$ magnitude maps at three 
  axial planes (left: $z = -5$~cm, middle: $z = 0$~cm and right: $z = 5$~cm.), while the bottom row shows the corresponding $B_1$ phase maps.  The dashed line indicates the coil winding surface.
  }
  \label{STS}
\end{figure}

\begin{figure}[htbp]
  \includegraphics[width=\textwidth]{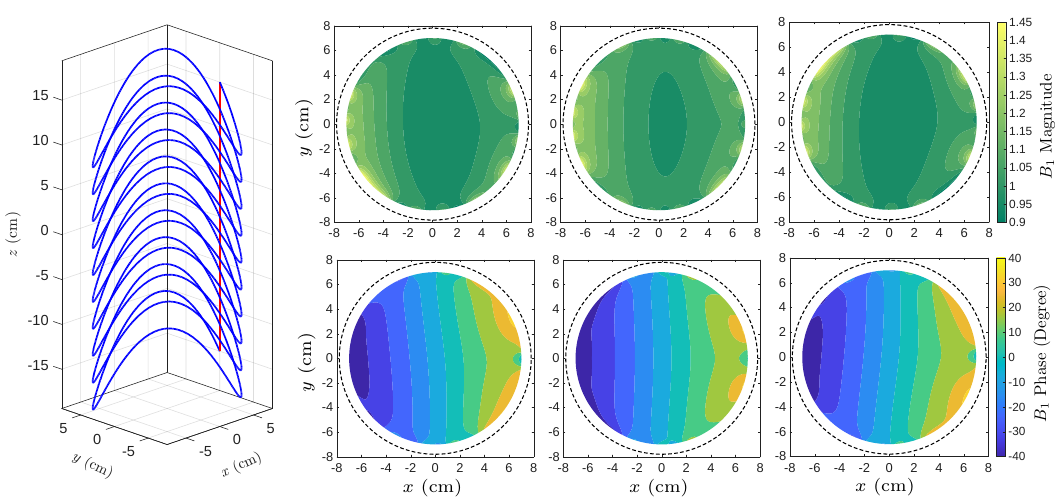}
  \captionof{figure}{Same as Figure~\ref{STS}, but for the RWTS coil.}
  \label{RWTS}
\end{figure}

\begin{figure}[htbp]
  \includegraphics[width=\textwidth]{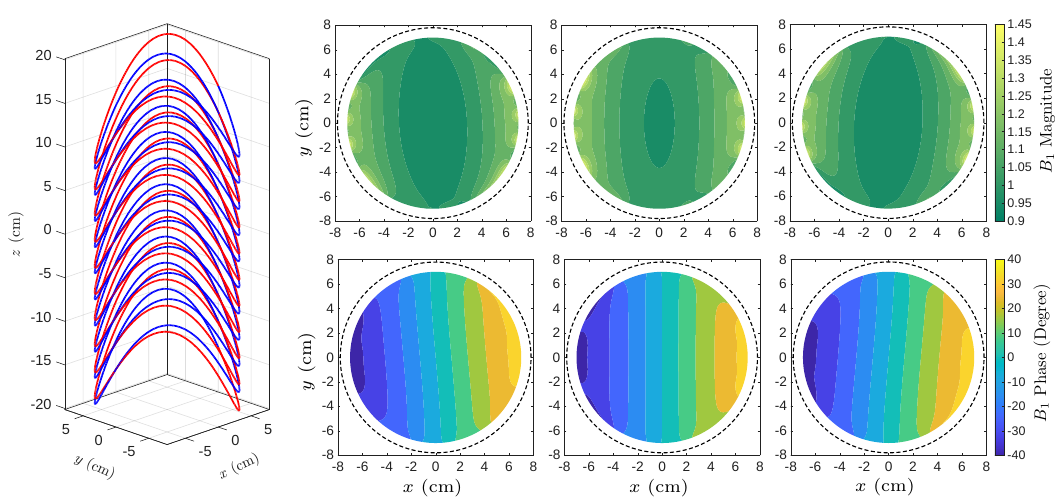}
  \captionof{figure}{Same as Figure~\ref{STS}, but for the DWTS coil.}
  \label{DWTS}
\end{figure}

\begin{figure}[htbp]
  \includegraphics[width=\textwidth]{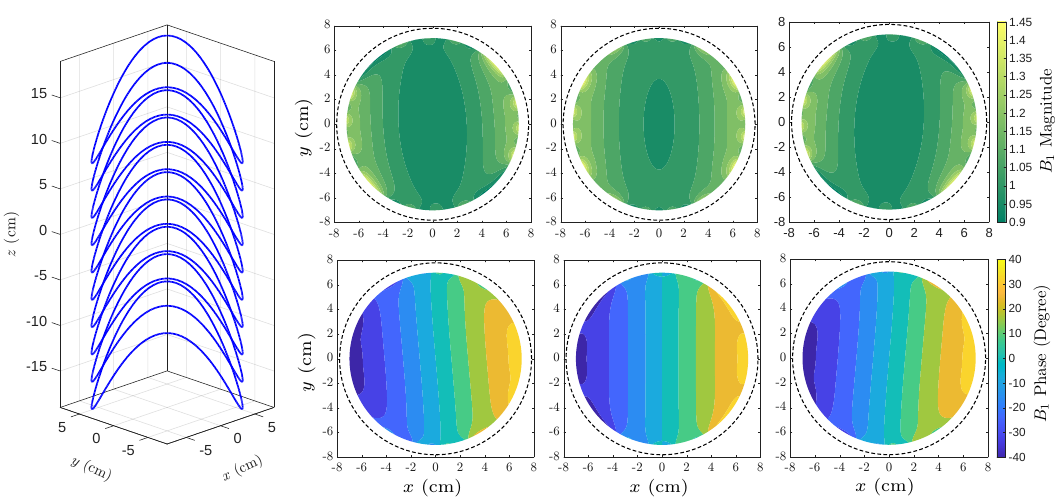}
  \captionof{figure}{Same as Figure~\ref{STS}, but for the DLTS coil.}
  \label{DLTS}
\end{figure}

\begin{figure}[htbp]
  \includegraphics[width=\textwidth]{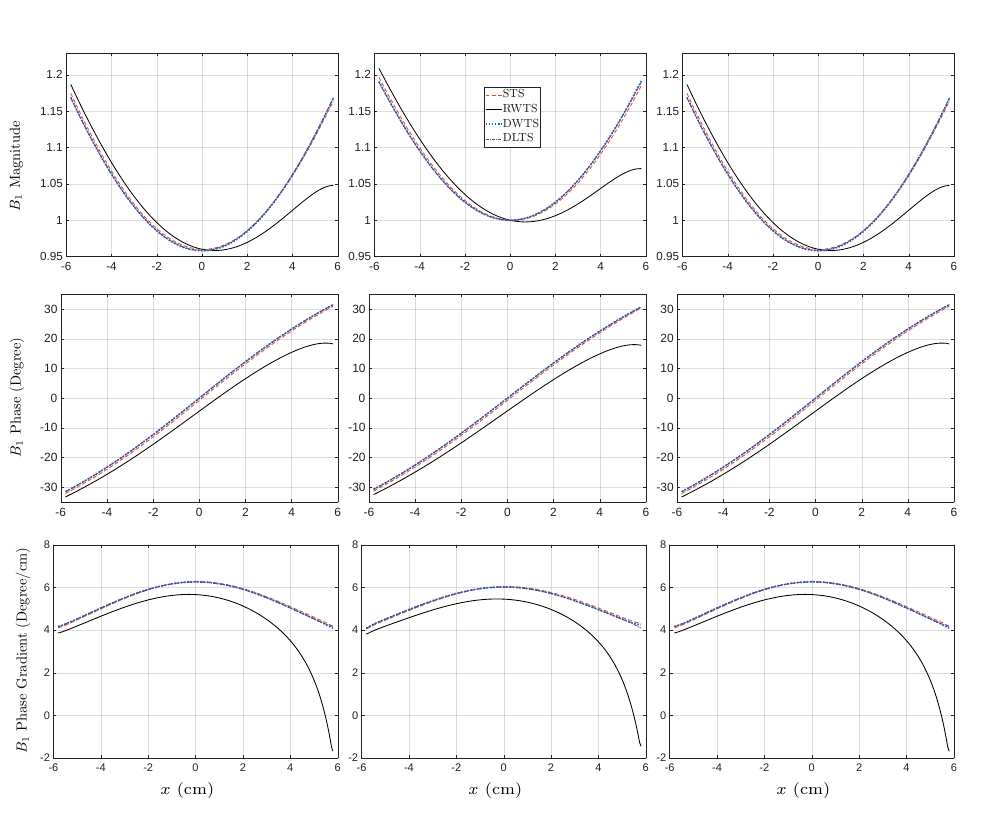}
 \caption{$B_1$ magnitude, phase, and phase gradient along the $x$-axis for $y=0$ in three axial planes $z = -5$ cm (left), $z = 0$ cm (middle) and $z = 5$ cm (right) across the four coil designs.
 The single legend refers to all panels.}
  \label{4coil_plots}
\end{figure}

The winding patterns explored here are shown in  Figs.~\ref{STS} through~\ref{DLTS}, along with contour plots of  $B_1$ magnitude and phase  at three axial planes ($z = -5$, 0, and 5~cm). The $B_1$ field magnitude  is normalized to its value at the coil isocenter $(0, 0, 0)$ for each coil type.  Figure~\ref{4coil_plots} provides a side-by-side comparison of  all coils for $B_1$ magnitude, phase, and phase gradient along  the $x-$axis at $y=0$ in each of the three aforementioned axial planes.  Table~\ref{tab:untruncated} provides a summary of the nominal characteristics of the four different coil types, including the mean phase gradient and total wire length.

As can be observed in Figs.~\ref{STS} through~\ref{4coil_plots}, the STS, DWTS and DLTS coils all produce very similar $B_1$ field distributions, with near equivalent phase and phase gradient profiles.  For the RWTS coil, on the other hand, the return wire introduces a strong asymmetry, resulting in significant degradation of all $B_1$ field characteristics.  Given that an STS coil cannot be realized in practice (since it lacks a return wire), the DWTS and DLTS designs emerge as excellent alternative winding patterns for twisted solenoid coils for  TRASE MRI, offering improved characteristics over the hitherto employed RWTS design.  Of these two, the DLTS offers the additional advantage of requiring much less wire to construct, as it does not need two layers.  Wire length can be an issue if it becomes comparable to the electromagnetic wavelength at the operating frequency~\cite{sunjmr}. (A rule of thumb is that it should be kept to less than $\lambda/20$~\cite{Hoult1978}.)  As can be seen in Table~\ref{tab:untruncated}, the DWTS coil requires nearly twice the length of wire as the DLTS coil. The DLTS coil itself only requires about 3\% more wire than the RWTS coil, however, making its substitution into existing and future TRASE MRI implementations an obvious and highly beneficial decision.   This is further supported by the fact the DLTS design provides the same desired central field and phase gradient strength produced by the idealized STS coil.

\begin{table}[H]
\caption{Summary of nominal characteristics of the four different coil types: the magnitude of $B_1$ at the coil isocenter, the mean phase gradient,  and the total wire length. 
The value of $\bar{G}_{x}$ is calculated along the $x-$axis at $y=z=0$ over the imaging volume diameter used in Ref.~\cite{sunjmr} (i.e.\ 70\% of the coil diameter or 10.9~cm).
For the DLTS, the total length also includes the connecting wires and return wire, which are not shown in Fig.~\ref{DLTS} for clarity. 
}
\begin{center}
\begin{tabular}{lccc}
Coil type & $|B_1(0,0,0)|$ ($\mu$T/A) & $\bar{G}_x$ (deg/cm) &Wire length (m)\\
\hline
STS & 36.61 & 5.38 &6.81\\
\hline
RWTS & 36.71& 4.50 &7.11\\
\hline
DWTS & 72.96 & 5.37 &13.62\\
\hline
DLTS & 36.52 & 5.36 &7.35\\
\end{tabular}
\end{center}
\label{tab:untruncated}
\end{table}

\newpage

\newpage

\subsection{Effect of shifting the angular location of the RWTS return wire}
As noted in the previous section, the RWTS coil exhibits the poorest performance in terms of $B_1$ magnitude uniformity and phase linearity. This degradation is due to the additional return wire segment, which significantly alters the RF magnetic field distribution, in particular the $B_{1y}$ component. It is worth noting, however, that the impact of this current segment can be mitigated to make the RWTS coil more suitable for TRASE MRI. 

To address this issue, the position of the return wire can be adjusted to minimize its contribution to $B_{1y}$, thereby improving the overall magnetic field performance of the coil.   This can be achieved by an appropriate combination of winding shift $\varphi$ and physical rotation of the coil about the $z$-axis to change the angular location of the return wire and preserve the desired phase gradient direction of the coil.
For clarity, and to avoid conflating the primary purpose of the winding shift, we propose a different, and in our opinion simpler, schema.
The specific steps are the following: (i) $\varphi$ is chosen as usual to set the phase gradient direction; (ii) an offset   $\theta_{\text{offset}}$  is added to the parameter $\theta$  of Eq.~\ref{parametric_curve3} to shift the start and end points of the coil, thereby relocating the return path; and (iii) the $B_1$ field of the coil is evaluated at, and relative to, its physical central plane, where imaging will occur.\footnote{The central plane is no longer at $z=0$ in the reference frame of Eq.~\ref{parametric_curve3}. In practice, the coil would be constructed and located in the MRI apparatus relative to its physical center, of course.}

\begin{figure}[htbp]
\centering
  \includegraphics[width=\textwidth]{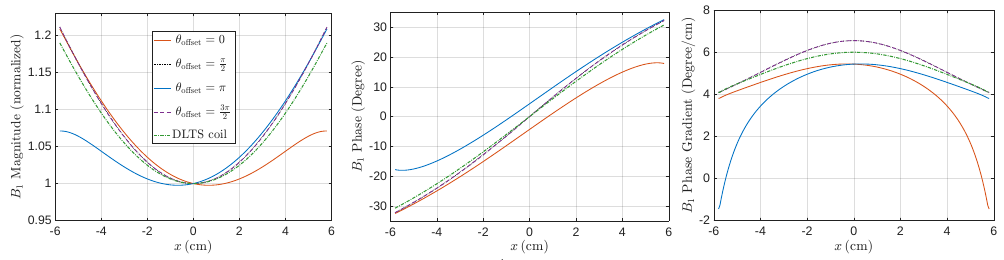}
 \caption{Comparison of the RWTS coil for four different offset angles and the DLTS coil: $B_1$ magnitude (left), phase (middle), and phase gradient (right) along the $x$-axis in the central plane. The single legend refers to all panels.}
  \label{offset}
\end{figure}

For the RWTS coils of interest here, the domain of Eq.~\ref{parametric_curve3} becomes $\theta~\epsilon~[-N \pi +\theta_{\text{offset}},~N \pi +\theta_{\text{offset}}]$, where $N$ remains the total number of turns.  To illustrate the effect of shifting the angular location of the RWTS return wire, we employ the same coil parameters of the previous section 
with $\theta_{\text{offset}} = 0, \pi/2, \pi,$ and $3\pi/2$.  
For $\theta_{\text{offset}} = 0$ and $\pi$, the return wire lies in the $y=0$ plane (at $x=+a$ and $-a$, respectively) and most strongly contributes to $B_{1y}$.   
For $\theta_{\text{offset}} = \pi/2$ and $3\pi/2$, the return wire lies in the $x=0$ plane (at $y=+a$ and $-a$, respectively) and most strongly contributes to $B_{1x}$, which does not factor into the definition of the transverse rf field $\bm B_1$ as given in Section~\ref{sec:B1eval}.  Results of $B_1$ evaluation in the central plane of each RWTS  coil are shown in Fig.~\ref{offset}, along with the those of the DLTS coil for comparison.  The results for the RWTS  coil with $\theta_{\text{offset}} = 0$ and the DLTS coil are the same as those presented in the central column of Fig.~\ref{4coil_plots}.

As can be seen in Fig.~\ref{offset}, moving the return wire to the $x=0$ plane, i.e.\  choosing $\theta_{\text{offset}} = \pi/2$ or $3\pi/2$,  greatly improves the $\bm B_1$ field characteristics of Eqs.~\ref{B1mag} through \ref{B1phaseGrad}.  The results also show that the DLTS coil is still the better choice across all metrics: $|B_1|$ homogeneity, $B_1$ phase linearity, and $G_x$ homogeneity.  It should also be noted again that the main RF field contribution of the improved RTWS coils is still present; it has just been shifted from $B_{1y}$ to  $B_{1x}$.   If $B_0 \gg B_{1x}$, the impact of this term is expected to be negligible~\cite{deng, sunjmr}. This would  certainly not be the case in very low $B_0$ field, however~\cite{bidinosti_artefacts}.   
Either way, the strong RF field, and its potentially  deleterious effects, generated by the return wire of an RTWS coil can be completely eliminated by switching to a DLTS  design.

\subsection{Comparison of truncated RWTS and DLTS}
\begin{figure}[htbp]
  \includegraphics[width=\textwidth]{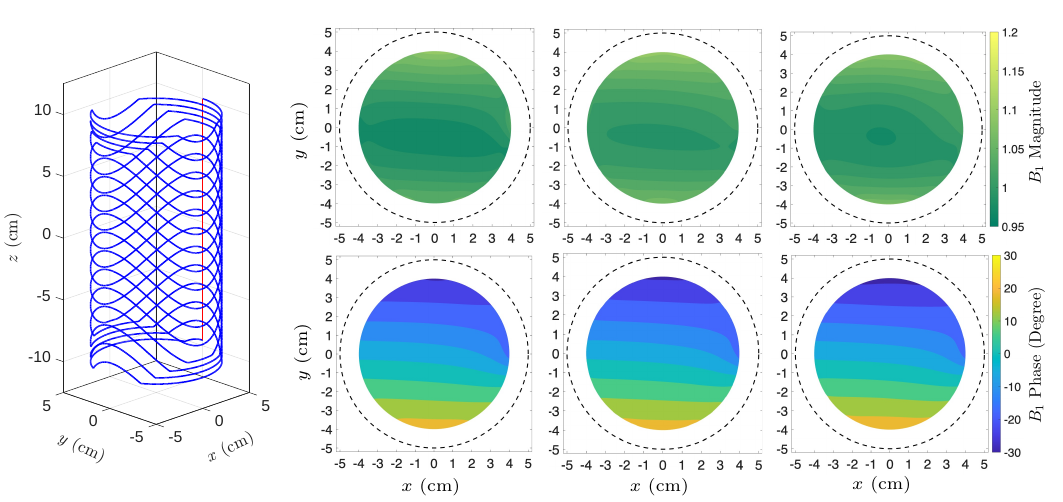}
  \captionof{figure}{
Results for the truncated version of the inner RWTS coil. The left plot shows the winding pattern. The top row of contour plots on the right displays $B_1$ magnitude maps at three 
  axial planes (left: $z = -5$~cm, middle: $z = 0$~cm and right: $z = 5$~cm.), while the bottom row shows the corresponding $B_1$ phase maps.  The dashed line indicates the coil winding surface.
}
  \label{RWTTS}
\end{figure}

\begin{figure}[htbp]
  \includegraphics[width=\textwidth]{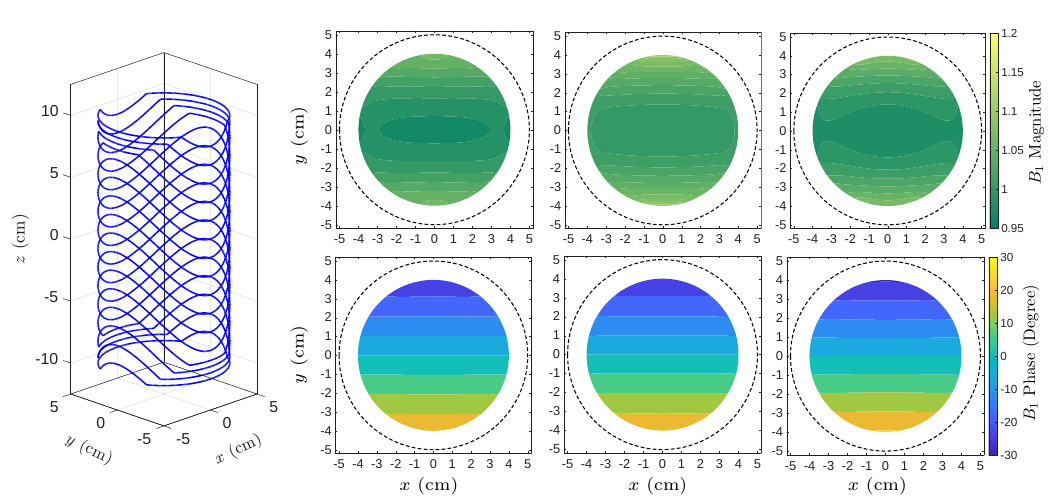}
  \captionof{figure}{Same as Figure~\ref{RWTTS}, but for the truncated inner DLTS coil.}
  \label{DLTTS}
\end{figure}

\begin{figure}[htbp]
  \includegraphics[width=\textwidth]{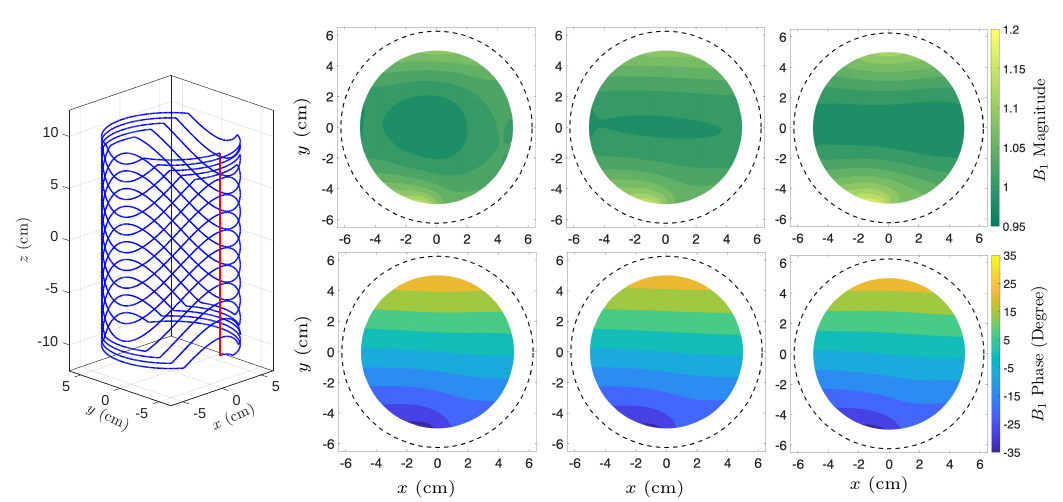}
  \captionof{figure}{Same as Figure~\ref{RWTTS}, but for the truncated outer RWTS coil.}
  \label{RWTTS_outer}
\end{figure}

\begin{figure}[htbp]
  \includegraphics[width=\textwidth]{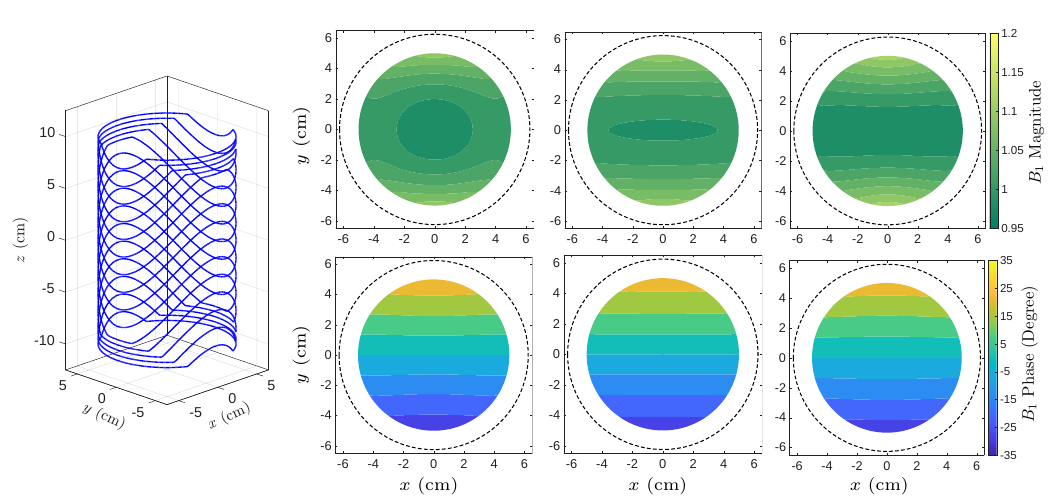}
  \captionof{figure}{Same as Figure~\ref{RWTTS}, but for the truncated outer DLTS coil.}
  \label{DLTTS_outer}
\end{figure}

To reduce the length of twisted solenoid coils while preserving phase gradient strength and field uniformity, a truncated coil design was introduced by Sedlock et al.~\cite{sedlock2023truncated}. This approach minimizes the $z$-coordinate of each point on the coil via the following formula:
\begin{equation}
P'(\theta) =
\begin{cases}
P(\theta) , & \text{if } |P(\theta)| \leq \frac{L}{2} - pg \\
\frac{L}{2} - pg, & \text{if } P(\theta) > \frac{L}{2} - pg \\
pg - \frac{L}{2}, & \text{if } P(\theta) < pg - \frac{L}{2}
\end{cases} \, ,
\label{truncatedPz}
\end{equation}
where$P(\theta)$ is from Eq.~\ref{parametric_curve3}, $L$ is the desired axial length of truncated coil, $g$ is the gap between the truncated end windings, and $p$  is a non-negative integer starting from zero, used to define where the truncation process considers the ``next turn'' to begin~\cite{sedlock2023truncated}.  The overall effect of Eq.~\ref{truncatedPz} is to flatten the outermost turns (or loops) of a twisted solenoid coil -- as can be seen in Figs~\ref{RWTTS} through~\ref{DLTTS_outer} --
thereby improving  usable imaging volume relative to  length.  The truncation process of Eq.~\ref{truncatedPz} is easily applied to our DLTS designs on a loop-by-loop basis, starting with $p=0$ for the outer most loops and working inward.

For the comparison in this section, we choose the \#1 inner and outer coils in Table~1 of Ref.~\cite{sedlock2023truncated}, which were the top ranking designs that came out of a parameter search study, subject to  selection criteria on the basis of $B_1$ field uniformity and magnitude, as well as phase gradient strength.  These coils were constructed and used as part of a decoupled, truncated coil pair for a successful demonstration of 1D TRASE MRI~\cite{sedlock2023truncated}.  Field mapping was performed and again showed close agreement with Biot-Savart simulations~\cite{sedlock2023truncated}.  The winding parameters for the inner coil are $N=16$, $a=5.0$~cm, $A =2.44$~cm, $\varphi =90^\circ$, and $h = 1.5$~cm. For the outer coil, they are $N=15$, $a=6.25$~cm, $A=3.32$~cm, $\varphi =90^\circ$, $h = 1.7$~cm.  Both were designed as $G_{-y}$ coils, but the outer coil was deployed with a physical rotation of $90^\circ$ about the $z$-axis, effectively turning it into a $G_{+y}$ coil (see Table~2 in Ref.~\cite{sunjmr}). The truncation parameters for both coils  are $L=20$~cm and $g=0.5$~cm .
The decoupling solenoids employed in Ref.~\cite{sedlock2023truncated} to reduce cross-talk between the inner and outer TRASE coils are not consider here.

Figures~\ref{RWTTS} through~\ref{DLTTS_outer} show contour maps of $B_1$ magnitude and phase  from Biot-Savart simulations of the RTWS and DLTS versions of the inner and outer truncated coils using the parameters listed above. The DLTS winding  pattern again provides  clear improvement in  field and phase uniformity over the standard RTWS design.  
There is negligible difference in the central field and phase gradient strength between corresponding versions of the coils (see Table~\ref{tab:truncated}). And with the DLTS requiring only about 3\% more wire to construct, there is no compromise in switching to this design.

\begin{table}[b]
\caption{Summary of nominal characteristics of the inner and outer truncated RWTS and DLTS coils: the magnitude of $B_1$ at the coil isocenter, the mean phase gradient,  and the total wire length. 
The value of $\bar{G}_{y}$ is calculated along the $y-$axis at $x=z=0$ over the imaging volume diameter used in Ref.~\cite{sedlock2023truncated} (i.e.\ 80\% of the inner coil diameter or 8~cm).
For the DLTS coils, the total length also includes the connecting wires and return wire, which are not shown in Figs.~\ref{DLTTS} and~\ref{DLTTS_outer} for clarity.
}
\begin{center}
\begin{tabular}{lccc}
Coil type & $|B_1(0,0,0)|$ ($\mu$T/A) & $\bar{G}_{y}$ (deg/cm) &Wire length (m)\\
\hline
RWTS inner & 78.57& -5.67 &6.02\\
\hline
DLTS inner & 78.38 & -5.75 & 6.21\\
\hline
\hline
RWTS outer & 68.44& 5.13 & 7.09\\
\hline
DLTS outer & 68.33 & 5.12 & 7.28\\
\end{tabular}
\end{center}
\label{tab:truncated}
\end{table}

\clearpage

\clearpage

\section{Conclusion}

Overall, this comparative study provides valuable information for improving the design of RF phase-gradient coils for TRASE MRI systems employing vertical $B_0$ magnets. 
The main goal of this work was the introduction of alternative winding patterns for twisted solenoid coils that have improved $B_1$ field uniformity and phase linearity  over existing designs.  The simple twisted solenoid (STS) has previously been used to guide TRASE coil design, but it fails to capture the deleterious effects of the return wire that must be present in any physical implementation of such a coil (i.e. an RWTS).  The magnitude of the impact of the return wire depends on its azimuthal location as well as the phase gradient direction 
of the coil.  In all cases, the unwanted RF field contributions of the return wire can be eliminated by switching to the proposed double-wound (DWTS) or discrete-loop twisted solenoid (DLTS) designs.  This was demonstrated through  Biot-Savart simulations using the design parameters of a variety of RWTS coils (e.g.\ $G_x$, $G_y$,  truncated, rotated) described in the literature.    The DLTS requires less wire to construct than the corresponding DWTS, however, making it the more practical choice of the two and less prone to potential 
 finite wavelength effects.  The DLTS also lends itself very well to truncated coil designs, and its design principles based on the stream function may be advantageous for future innovation.  Given that  3D printed coil formers are presently employed for making twisted solenoid coils for  TRASE MRI, a switch to either the DWTS or DLTS design should be very straightforward and  beneficial.  
 Lastly, we also presented a novel schema for shifting the location of the return wire in an RWTS coil, should one choose to stick with that design.

 \section*{CRediT authorship contribution statement}
\noindent \textbf{Nahid Ghomimolkar:}
Methodology, Software, Validation, Formal analysis,  Investigation,  Writing -- original draft, Visualization.
 \textbf{Alexander E.\ Krosney:}
Methodology, Software, Validation, Formal analysis,  Investigation,  Writing -- original draft, Visualization.
 \textbf{Christopher P.\ Bidinosti:}
 Conceptualization, Methodology,   Validation , Formal analysis,  Investigation,  Writing -- original draft,  Writing -- review \& editing, Visualization, Supervision, Project administration, Funding acquisition.
 
\section*{Declaration of Competing Interest}
The authors declare that they have no known competing financial interests or personal relationships that could have appeared
to influence the work reported in this paper.

\section*{Acknowledgments}
We gratefully acknowledge the support of the University of
Winnipeg, the University of Manitoba, and the Natural Sciences and Engineering Research Council of Canada [grant number 2020-06191], especially  the Undergraduate Student Research Award for AEK.

\bibliographystyle{unsrt}
\bibliography{AWP}

@article{sharp,
author = {Sharp, J.C. and King, S.B.},
title = {{MRI} using radiofrequency magnetic field phase gradients},
journal = {Magnetic Resonance in Medicine},
volume = {63},
number = {1},
pages = {151-161},
doi = {10.1002/mrm.22188},
year = {2010}
}

@article{sharp2,
author = {Sharp, J.C. and King, S.B. and Deng, Q. and Volotovskyy, V. and Tomanek, B.},
title = {High-resolution {MRI} encoding using radiofrequency phase gradients},
journal = {NMR in Biomedicine},
volume = {26},
number = {11},
pages = {1602-1607},
doi = {10.1002/nbm.3023},
year = {2013}
}

@article{bidinosti_artefacts,
title = {Concomitant {B1} Field in Low-Field {MRI}: Potential Contributions to {TRASE} Image Artefacts},
author = {Bidinosti, C.P. and Nacher, P.-J. and Tastevin, G.},
url = {https://hal.archives-ouvertes.fr/hal-02366411},
booktitle = {Joint annual meeting ISMRM-ESMRMB 2018},
address = {Paris, France},
organization = {Proc. of the Intl. Soc. Mag. Reson. Med.},
journal = {Proc. of the Intl. Soc. Mag. Reson. Med.},
volume = {26},
pages = {2670},
year = {2018},
month = Jun,
hal_id = {hal-02366411},
hal_version = {v1}
}

@article{kumaragamage,
title = {{B1} phase gradient coil design for low field exploration of {TRASE} {MRI}},
author = {Kumaragamage, S. and Lang, M. and Ostapchuk, D. and Bidinosti, C.P.},
journal = {Proc. of the ESMRMB 33, Magn. Reson. Mater. Phy.},
year = {2016},
volume = {29},
pages = {S34},
booktitle = {}
}

@article{bellec138,
title = {Target Field Based {RF} Phase Gradient Transmit Array for {3D} {TRASE} {MRI}},
author = {Bellec, J. and King, S.B. and Liu, C.-Y. and Bidinosti, C.P.},
journal = {Proc. of the Intl. Soc. Mag. Reson. Med.},
year = {2013},
volume = {21},
pages = {138},
booktitle = {}
}

@article{bellec723,
title = {A Target Field Approach to the Design of {RF} Phase-Gradient Coils},
author = {Bellec, J. and Liu, C.-Y. and King, S.B. and Bidinosti, C.P.},
journal = {Proc. of the Intl. Soc. Mag. Reson. Med.},
year = {2011},
volume = {19},
pages = {723},
booktitle = {}
}

@article{bohidar,
title = {{TRASE} {1D} sequence performance in imperfect ${B}_1$ fields},
journal = {Journal of Magnetic Resonance},
volume = {305},
pages = {77 - 88},
year = {2019},
issn = {1090-7807},
doi = {10.1016/j.jmr.2019.06.005},
author = {Bohidar, P. and Sun, H. and Sarty, G.E. and Sharp, J.C.}
}

@article{nacher,
title = {A fast {MOSFET} rf switch for low-field {NMR} and {MRI}},
journal = {Journal of Magnetic Resonance},
volume = {310},
pages = {106638},
year = {2020},
issn = {1090-7807},
doi = {10.1016/j.jmr.2019.106638},
author = {Nacher, P.-J. and Kumaragamage, S. and Tastevin, G. and Bidinosti, C.P.}
}

@article{deng,
author = {Deng, Q. and King, S.B. and Volotovskyy, V. and Tomanek, B. and Sharp, J.C.},
year = {2013},
month = {04},
pages = {},
title = {{B}$_1$ transmit phase gradient coil for single-axis {TRASE} {RF} encoding},
volume = {31},
journal = {Magnetic resonance imaging},
doi = {10.1016/j.mri.2013.03.017}
}

@article{stockmann,
author = {Stockmann, J.P. and Cooley, C.Z. and Guerin, B. and Rosen, M.S. and Wald, L.L.},
year = {2016},
pages = {36--48},
title = {Transmit Array Spatial Encoding ({TRASE}) using broadband {WURST} pulses for {RF} spatial encoding in inhomogeneous {B0} fields},
volume = {268},
journal = {Journal of Magnetic Resonance},
doi = {10.1016/j.jmr.2016.04.005}
}

@article{sunjmr,
title = {The twisted solenoid {RF} phase gradient transmit coil for {TRASE} imaging},
journal = {Journal of Magnetic Resonance},
volume = {299},
pages = {135 - 150},
year = {2019},
issn = {1090-7807},
doi = {10.1016/j.jmr.2018.12.015},
author = {Sun, H. and Yong, S. and Sharp, J.C.}
}

@article{sunmrm,
author = {Sun, H. and Al-Zubaidi, A. and Purchase, A. and Sharp, J.C.},
title = {A geometrically decoupled, twisted solenoid single-axis gradient coil set for {TRASE}},
journal = {Magnetic Resonance in Medicine},
volume = {83},
number = {4},
pages = {1484-1498},
doi = {10.1002/mrm.28003},
year = {2020}
}

@article{queval,
author={Quéval, L. and Gottkehaskamp, R.},
journal={IEEE Transactions on Applied Superconductivity}, 
title={Analytical Field Calculation of Modulated Double Helical Coils}, 
year={2015},
volume={25},
number={6},
pages={1-7},
doi={10.1109/TASC.2015.2477377}
}

@article{alonso,
title = {Double helix dipole design applied to magnetic resonance: A novel {NMR} coil},
journal = {Journal of Magnetic Resonance},
volume = {235},
pages = {32 - 41},
year = {2013},
issn = {1090-7807},
doi = {10.1016/j.jmr.2013.07.004},
author = {Alonso, J. and Soleilhavoup, A. and Wong, A. and Guiga, A. and Sakellariou, D.}
}

@article{goodzeit,
author = {Goodzeit, C.L. and Ball, M.J. and Meinke, R.B.},
journal = {IEEE Transactions on Applied Superconductivity}, 
title = {The double-helix dipole - a novel approach to accelerator magnet design}, 
year = {2003},
volume = {13},
number = {2},
pages = {1365-1368},
doi = {10.1109/TASC.2003.812672}
}

@article{brideson,
author = {Brideson, M.A. and Forbes, L.K. and Crozier, S.},
title = {Determining complicated winding patterns for shim coils using stream functions and the target-field method},
journal = {Concepts in Magnetic Resonance},
volume = {14},
number = {1},
pages = {9-18},
doi = {10.1002/cmr.10000},
year = {2002}
}

@article{bidinosti_saddle,
title = {Active shielding of cylindrical saddle-shaped coils: Application to wire-wound {RF} coils for very low field {NMR} and {MRI}},
journal = {Journal of Magnetic Resonance},
volume = {177},
number = {1},
pages = {31 - 43},
year = {2005},
issn = {1090-7807},
doi = {10.1016/j.jmr.2005.07.003},
author = {Bidinosti, C.P. and Kravchuk, I.S. and Hayden, M.E.}
}

@article{peeren,
title = {Stream function approach for determining optimal surface currents},
journal = {Journal of Computational Physics},
volume = {191},
number = {1},
pages = {305 - 321},
year = {2003},
issn = {0021-9991},
doi = {10.1016/S0021-9991(03)00320-6},
author = {Peeren, G.N.}
}

@article{meinke2003modulated,
  title={Modulated double-helix quadrupole magnets},
  author={Meinke, RB and Goodzeit, CL and Ball, MJ},
  journal={IEEE transactions on applied superconductivity},
  volume={13},
  number={2},
  pages={1369--1372},
  year={2003},
  publisher={IEEE}
}

@article{hanson2002compact,
  title={Compact expressions for the {Biot-Savart} fields of a filamentary segment},
  author={Hanson, James D and Hirshman, Steven P},
  journal={Physics of Plasmas},
  volume={9},
  number={10},
  pages={4410--4412},
  year={2002},
  publisher={Citeseer}
}

@article{sedlock2023truncated,
  title={A truncated twisted solenoid {RF} phase gradient transmit coil for {TRASE} {MRI}},
  author={Sedlock, Christopher J and Purchase, Aaron R and Tomanek, Boguslaw and Sharp, Jonathan C},
  journal={Journal of Magnetic Resonance},
  volume={347},
  pages={107361},
  year={2023},
  publisher={Elsevier}
}

@Article{Sedlock2025,
  author    = {Sedlock, Christopher J. and Purchase, Aaron R. and Tomanek, Boguslaw and Sharp, Jonathan C.},
  journal   = {Magnetic Resonance in Medicine},
  title     = {Radial {TRASE}: {2D} {RF} encoding through mechanical rotation and active digital decoupling},
  year      = {2025},
  issn      = {1522-2594},
  month     = oct,
  doi       = {10.1002/mrm.70104},
  publisher = {Wiley},
}

@Article{Sarty2025,
  author    = {Sarty, Gordon E. and Vidarsson, Logi and Hansen, Christopher and Corrigal, Keifer and Sutherland, Lionel and Jamieson, Millie and Hogue, Micheal and Kassahun, Haile and Greyeyes, William and Teixeira, David and Goertzen, Lawrence and McEvoy, Jonathan and Pollard, Mark},
  journal   = {Frontiers in Neuroimaging},
  title     = {Learning to build low-field {MRI}s for remote northern communities},
  year      = {2025},
  issn      = {2813-1193},
  month     = jan,
  volume    = {3},
  doi       = {10.3389/fnimg.2024.1521517},
  publisher = {Frontiers Media SA},
}

@Article{Bohidar2020,
  author    = {Bohidar, Pallavi and Sun, Hongwei and Sharp, Jonathan C. and Sarty, Gordon E.},
  journal   = {Magnetic Resonance Imaging},
  title     = {The effects of coupled {B1} fields in {B1} encoded {TRASE} {MRI} - A simulation study},
  year      = {2020},
  issn      = {0730-725X},
  month     = dec,
  pages     = {74--83},
  volume    = {74},
  doi       = {10.1016/j.mri.2020.09.003},
  publisher = {Elsevier BV},
}

@Article{Sarty2021,
  author    = {Sarty, Gordon E.},
  journal   = {Linear Algebra and its Applications},
  title     = {Natural reconstruction coordinates for imperfect {TRASE} {MRI}},
  year      = {2021},
  issn      = {0024-3795},
  month     = feb,
  pages     = {94--117},
  volume    = {611},
  doi       = {10.1016/j.laa.2020.11.022},
  publisher = {Elsevier BV},
}

@Article{Der2018,
  author    = {Der, Eric and Volotovskyy, Vyacheslav and Sun, Hongwei and Tomanek, Boguslaw and Sharp, Jonathan C.},
  journal   = {Concepts in Magnetic Resonance Part B: Magnetic Resonance Engineering},
  title     = {Design of a high power {PIN}‐diode controlled switchable {RF} transmit array for {TRASE} {RF} imaging},
  year      = {2018},
  issn      = {1552-504X},
  month     = feb,
  number    = {1},
  volume    = {48B},
  doi       = {10.1002/cmr.b.21365},
  publisher = {Wiley},
}

@Article{Purchase2019,
  author    = {Purchase, Aaron R. and Pałasz, Tadeusz and Sun, Hongwei and Sharp, Jonathan C. and Tomanek, Boguslaw},
  journal   = {Magnetic Resonance Materials in Physics, Biology and Medicine},
  title     = {A high duty-cycle, multi-channel, power amplifier for high-resolution radiofrequency encoded magnetic resonance imaging},
  year      = {2019},
  issn      = {1352-8661},
  month     = jun,
  number    = {6},
  pages     = {679--692},
  volume    = {32},
  doi       = {10.1007/s10334-019-00763-1},
  publisher = {Springer Science and Business Media LLC},
}

@Article{Purchase2021,
  author    = {Purchase, Aaron R. and Vidarsson, Logi and Wachowicz, Keith and Liszkowski, Piotr and Sun, Hongwei and Sarty, Gordon E. and Sharp, Jonathan C. and Tomanek, Boguslaw},
  journal   = {IEEE Access},
  title     = {A Short and Light, Sparse Dipolar {Halbach} Magnet for {MRI}},
  year      = {2021},
  issn      = {2169-3536},
  pages     = {95294--95303},
  volume    = {9},
  doi       = {10.1109/access.2021.3093530},
  publisher = {Institute of Electrical and Electronics Engineers (IEEE)},
}

@Article{Bidinosti2022,
  author    = {Bidinosti, Christopher P. and Tastevin, Geneviève and Nacher, Pierre-Jean},
  journal   = {Journal of Magnetic Resonance},
  title     = {Generating accurate tip angles for {NMR} outside the rotating-wave approximation},
  year      = {2022},
  issn      = {1090-7807},
  month     = dec,
  pages     = {107306},
  volume    = {345},
  doi       = {10.1016/j.jmr.2022.107306},
  publisher = {Elsevier BV},
}

@Article{Hoult1978,
  author    = {Hoult, D.I.},
  journal   = {Progress in Nuclear Magnetic Resonance Spectroscopy},
  title     = {The {NMR} receiver: A description and analysis of design},
  year      = {1978},
  issn      = {0079-6565},
  month     = jan,
  number    = {1},
  pages     = {41--77},
  volume    = {12},
  doi       = {10.1016/0079-6565(78)80002-8},
  publisher = {Elsevier BV},
}
\end{document}